\documentclass[a4paper]{jpconf}
\usepackage{graphicx}
\begin{document}
\title{The potential of discrimination methods in a high pressure xenon TPC
for the search of the neutrinoless double-beta decay of Xe-136}
\author{F.J.~Iguaz, F.~Aznar\footnote{Present Address:
Centro Universitario de la Defensa, Universidad de Zaragoza, Spain.},
J.F.~Castel, S.~Cebri\'an, T.~Dafni, J.~Gal\'an, J.G.~Garza,
I.G.~Irastorza, G.~Luz\'on, H.~Mirallas and E.~Ruiz-Choliz}
\address{Laboratorio de F\'isica Nuclear y Astropart\'iculas, Universidad de Zaragoza, Spain.}
\ead{iguaz@unizar.es}

\begin{abstract}
In the search for the neutrinoless double beta decay of $^{136}$Xe, a high pressure xenon time projection
chamber (HPXe-TPC) has two advantages over liquid xenon TPCs: a better energy resolution and the
access to topological features, which may provide extra discrimination from background events.
%, keeping good signal efficiency.
The PandaX-III experiment has recently proposed a 200 kg HPXe-TPC based on Micromegas
readout planes, to be located at the Jinping Underground Laboratory in China. Its detection concept is
based on two results obtained within the T-REX project: Micromegas readouts can be built with
extremely low levels of radioactivity; and the operation in xenon-trimethylamine at 10 bar
in realistic experimental conditions has proven an energy resolution of 3\%~FWHM at the region of interest.
In this work, two discrimination methods are applied to
simulated signal and background data in a generic
200 kg HPXe-TPC, based on two well-known algorithms of graph theory:
the identification of connections and the search for the longest path.
Rejection factors greater than 100 are obtained for small pixel sizes and a signal efficiency of 40\%.
Moreover, a new observable (the blob charge density) rejects better surface contaminations,
which makes the use of a trigger signal ($T_0$) not imperative in this experiment.
\end{abstract}

\section{Introduction}
Experiments based on liquid xenon are leading the search of neutrinoless double beta decay of $^{136}$Xe
due to the fast development of this detection technique in the last decade~\cite{Gando:2016ag}.
More recently, experiments based on high pressure xenon
time projection chambers (HPXe-TPCs)~\cite{GomezCadenas:2014jjgc, Galan:2016jg}
have been proposed due to their better intrinsic energy resolution ($\sim$1\%~FWHM)
and the access to topological features,
which may provide extra discrimination from background events, keeping good signal efficiency.
In this context, the use of Micromegas charge readout planes~\cite{Andriamonje:2010sa} in a HPXe-TPC
has been studied by T-REX project\footnote{T-REX webpage: http://gifna.unizar.es/trex},
leading to the following results:
Micromegas readouts show extremely low levels of radioactivity
(below 0.1~$\mu$Bq/cm$^2$ for both $^{214}$Bi and $^{208}$Tl)~\cite{Irastorza:2016ii};
they have been operated in xenon-trimethylamine (Xe-TMA) at 10 bar
in realistic experimental conditions (30~cm diameter readout, 1200~channels, 38~cm drift),
proving an energy resolution of 3\%~FWHM at the Range of Interest (RoI)~\cite{Gonzalez:2016dg};
%still limited by practical and not fundamental reasons;
and the combination of high granularity readout planes
and low diffusion (as measured in Xe-TMA) can reduce the background level by two-three orders of magnitude
keeping a signal efficiency of 40\%~\cite{Cebrian:2013sc}.

Micromegas readout planes will be used in the first module of PandaX-III experiment~\cite{Galan:2016jg},
to be installed at the China Jinping Underground Laboratory (CJPL)~\cite{Li:2015jl} by 2017.
A half-life sensitivity of $10^{26}$~y (at 90\% CL) is expected after 3 years of live-time,
supposing a~3\%~FWHM energy resolution at Q-value (2458 keV),
a background level of $\sim 10^{-4}$ counts keV$^{-1}$kg$^{-1}$y$^{-1}$ at the RoI
and a signal efficiency of~35\%.
The potential of discrimination methods in this detection technique,
which combines a low diffusion gas like Xe-TMA and readout planes with pixel sizes of $\sim$1~mm,
is the aim of this paper and will be further developed in a future publication.

\section{Simulation and discrimination algorithms}
\begin{figure}[htb!]
\centering
\includegraphics[width=75mm]{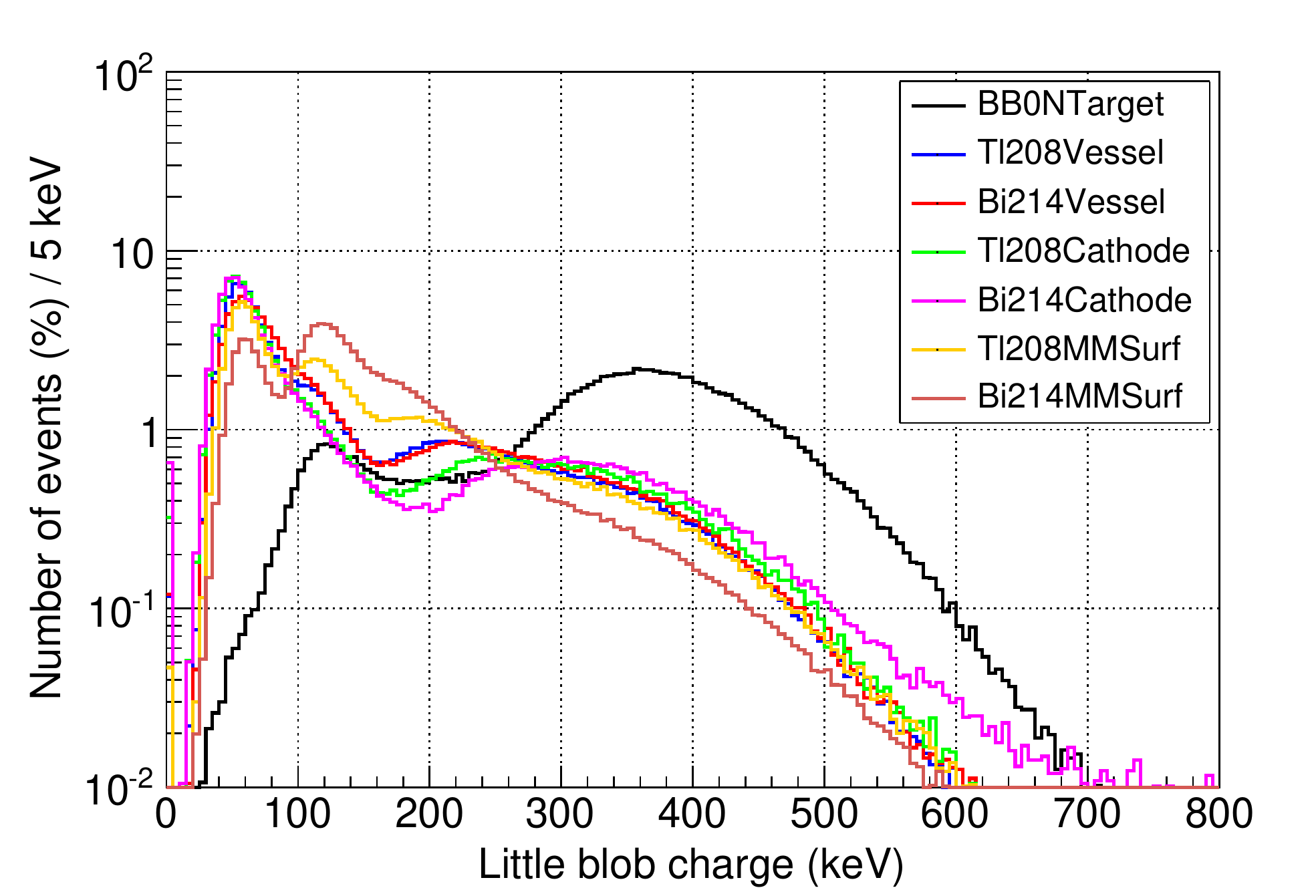}
\includegraphics[width=75mm]{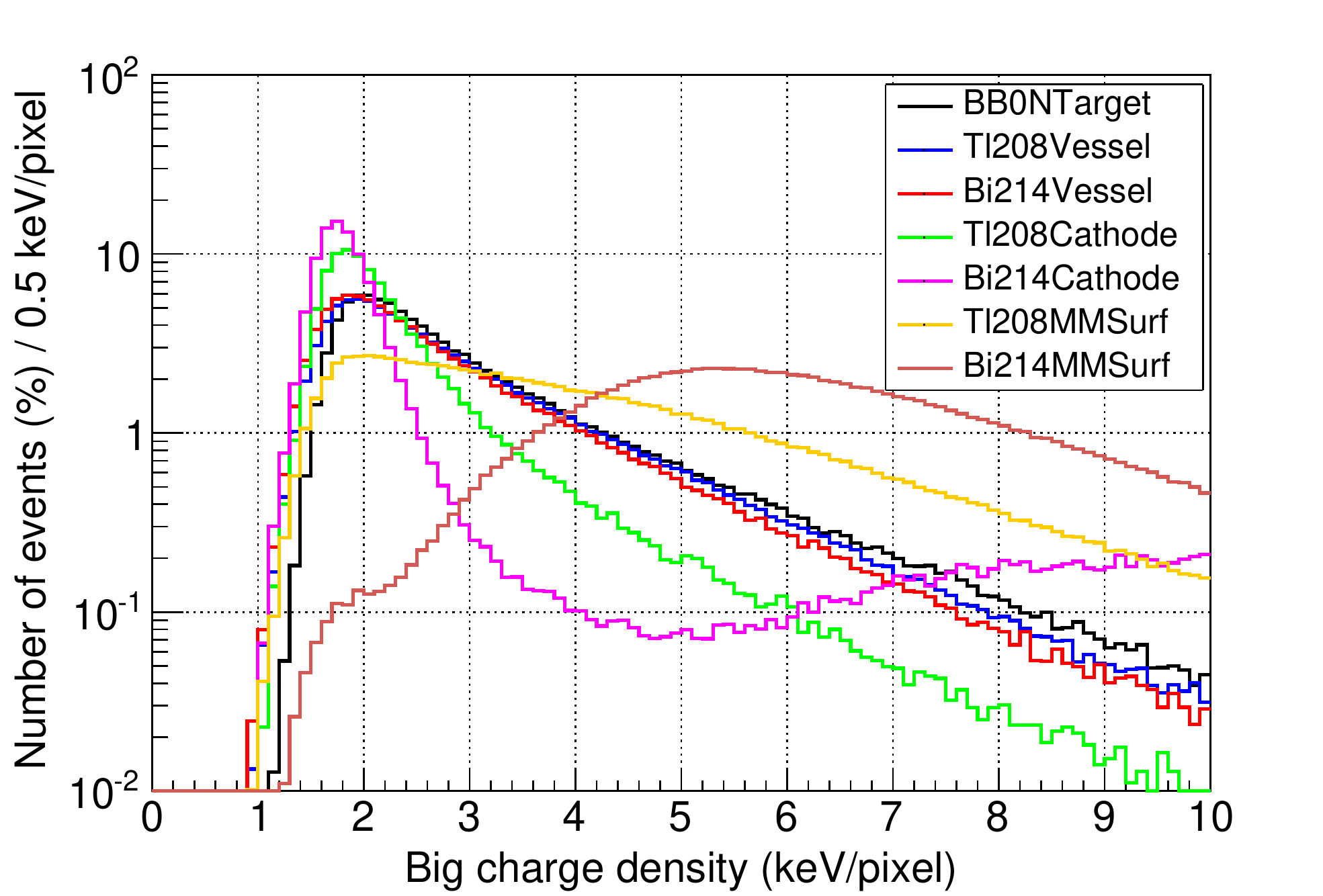}
\caption{\it Distribution of the little blob charge (left) and the big blob density (right)
in a 200~kg HPXeTPC for signal events (black line) and different contaminations:
vessel (blue and red lines for $^{208}$Tl and $^{214}$Bi, respectively),
cathode (green and magenta lines) and Micromegas readout planes (yellow and brown lines).
Events are in the RoI and are single-track.}
\label{fig:TopRejection}
\end{figure}

A 200~kg HPXeTPC filled with Xe+1\%TMA at 10 bar has been simulated by Geant4 and REST codes~\cite{Iguaz:2015fji},
fixing a pixel size of 2~mm and an energy resolution of 1\%~FWHM at the RoI.
In the analysis, tracks\footnote{Group of points for which we can always find a continuous path between any pair of points.}
and blobs\footnote{A big energy deposit at the end of one electron path.} are found by two well-known algorithms
of graph theory: the identification of connections (or tracks) and the search for the longest path.
Signal events are then selected by three discrimination criteria: a fiducial area; a single-track condition,
with some possible bremsstrahlung photons (energy below 40~keV) situated near the main track (12~cm maximum);
and some limits in the blob charge and density.
The distribution of these two last observables are shown in in Fig.~\ref{fig:TopRejection}
to illustrate their discrimination power.
The little blob charge distribution (black line) is quite separated from background distributions,
as already shown by the Gotthard experiment~\cite{Wong:1993uq}.
Meanwhile, the blob density mainly rejects surface contaminations at the readout planes
(yellow and brown lines), as they are deposited near the readouts, drift very short distances
and show bigger charge densities.

\begin{figure}[htb!]
\centering
\includegraphics[width=75mm]{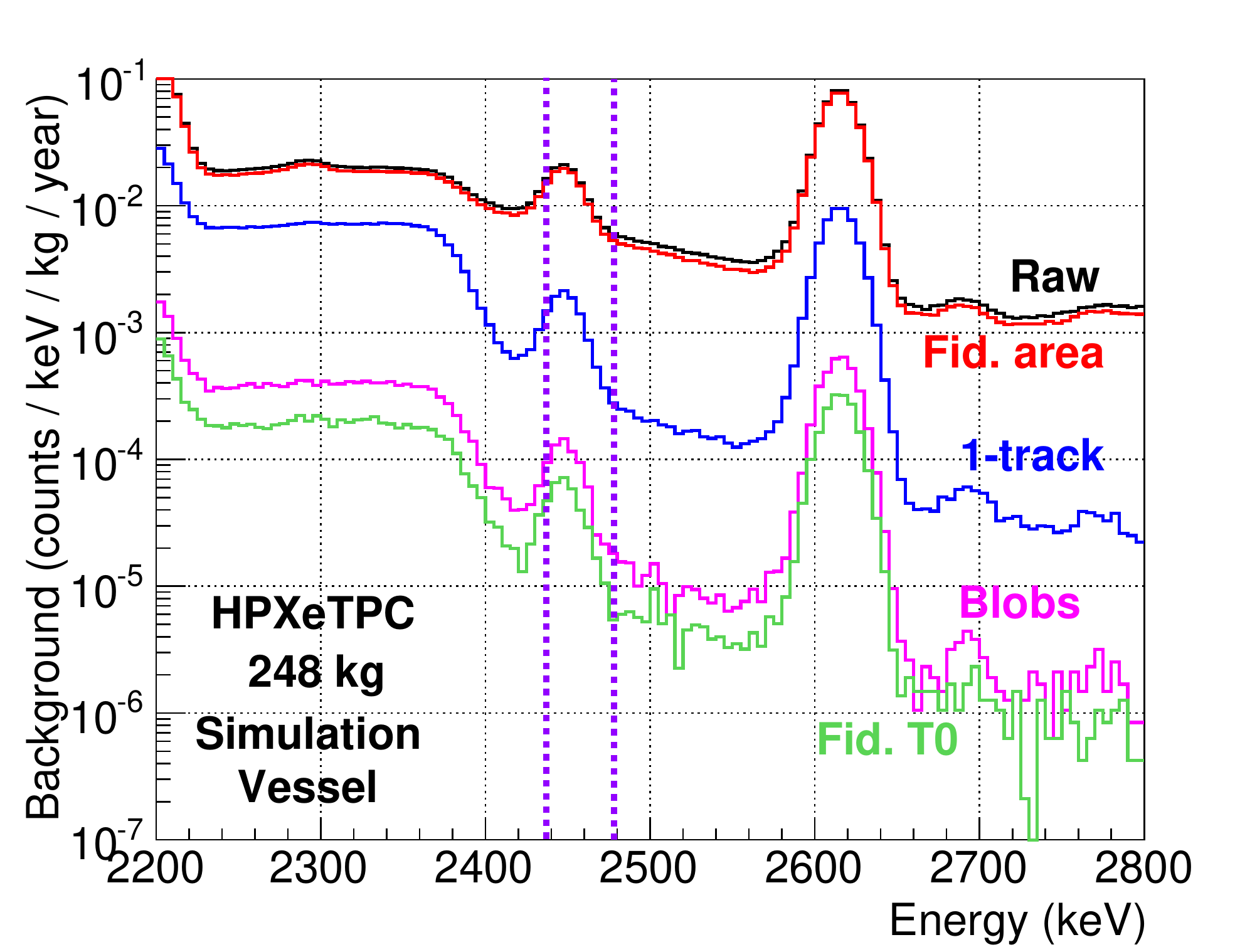}
\includegraphics[width=75mm]{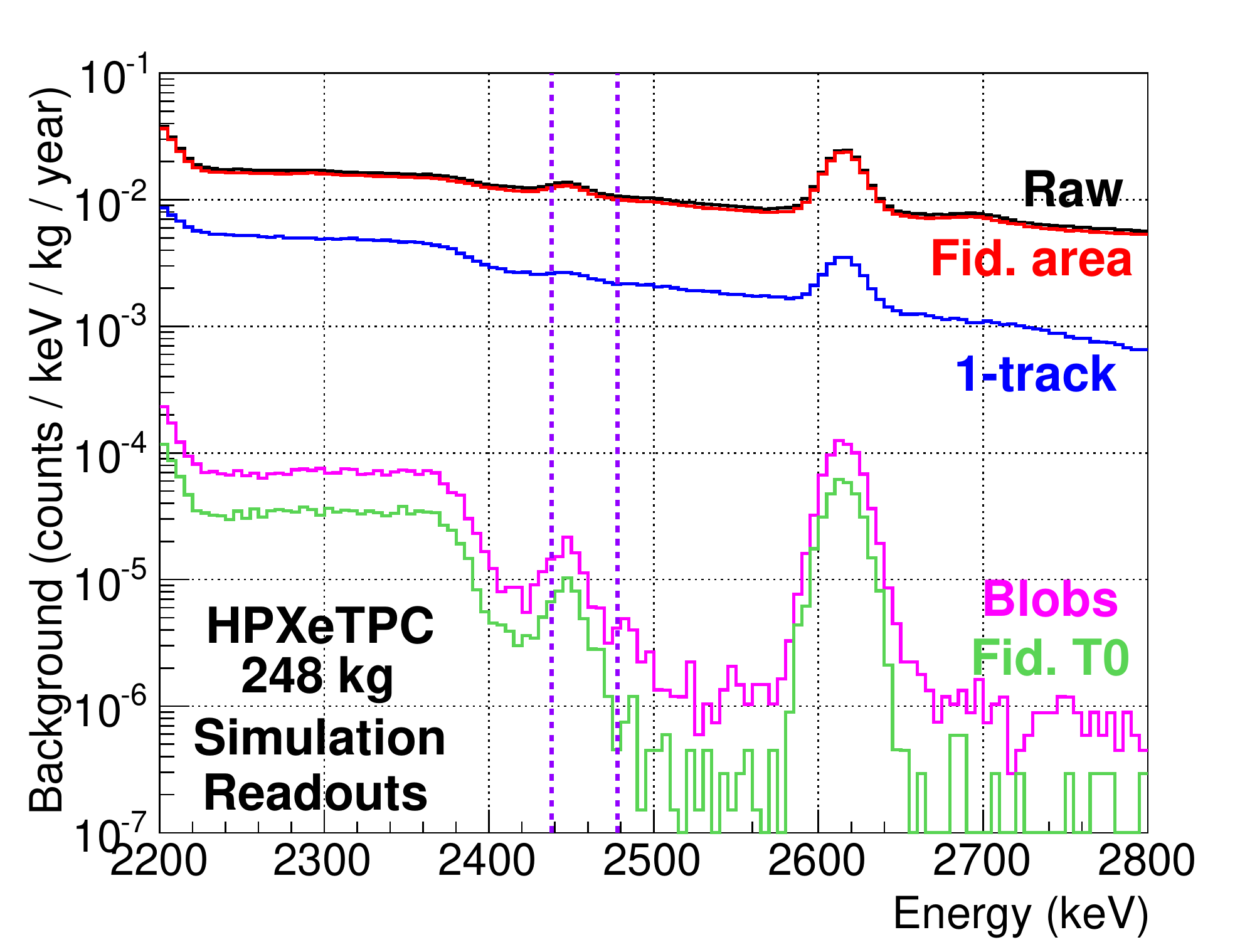}
\caption{\it Energy spectra generated by $^{208}$Tl and $^{214}$Bi events emitted
from the vessel (left) and the Micromegas readout planes (right)
in 200~kg HPXeTPC after the successive application of the selection criteria: raw data (black line),
fiducial area (red line), single-track (blue line), blob observables (magenta line)
and a trigger signal or T$_0$ (green line).
The vertical violet dotted lines delimit a RoI of 40 keV, equivalent to a 1\%~FWHM energy resolution.}
\label{fig:TopSpectra}
\end{figure}

Fixing a signal efficiency\footnote{This signal efficiency includes the intrinsic and the analysis efficiencies.} of 40\%
and scaling the results by the material radioactivity activities in~\cite{Irastorza:2016ii},
the background energy spectra of Fig.~\ref{fig:TopSpectra} are obtained for the vessel and readout planes.
The background level of these components in the RoI is reduced by two and three orders of magnitude respectively,
down to $(8.1 \pm 0.2)$ and $(1.3 \pm 0.1) \times 10^{-5}$ counts keV$^{-1}$kg$^{-1}$y$^{-1}$.
The blob criteria reject a factor $> 15$ of background events for the vessel,
a value better than Gotthard experiment~\cite{Wong:1993uq},
and a factor $> 55$ for the readouts.
The absence of a trigger signal (or $T_0$) by registering the primary scintillation
only gives a factor $\sim 2$ more background events in the RoI:
this indicates that the implementation of a $T_0$ -unwanted due to the technological
and radiopurity cost of a light readout- is not imperative.

\section{Prospects}
These motivating results cannot be directly translated to a background model of a HPXeTPC
because the electronics and the readout planes must be included in the simulation, as done in~\cite{Iguaz:2015fji}.
Then either the algorithms in~\cite{Cebrian:2013sc} are adopted to
the two 2D views for each event, as done in~\cite{Wong:1993uq},
or the event's 3D track is previously reconstructed from these two views.
This second option has been extensively studied by liquid argon detectors,
like ICARUS~\cite{Antonello:2013ma} and DUNE~\cite{Marshall:2015jsm}.
However, its application to $0\nu\beta\beta$ is a challenging task
as MeV electron tracks are relatively short compared to their complex topology 
and there are only two 2D views for each event.

\ack
We acknowledge the support from the European Commission under the European Research Council
T-REX Starting Grant ref. ERC-2009-StG-240054 of the IDEAS program of the 7th EU Framework Program.
F.I. and T.D. acknowledge the support from the \emph{Juan de la Cierva} and \emph{Ram\'on y Cajal} programs
of the Spanish Ministry of Economy and Competitiveness.

\end{document}